\documentclass[aps,prb,preprint,showkeys,showpacs]{revtex4-2}
\pdfoutput=1
\usepackage{amsmath}
\usepackage{amssymb}
\usepackage{amsfonts}
\usepackage{graphicx}
\usepackage{bm}
\usepackage{siunitx}
\usepackage{color}
\usepackage{hyperref}
\hypersetup{%
    pdfborder = {0 0 0}
}

\allowdisplaybreaks
\usepackage{cleveref}

\graphicspath{{./figures/}}

\renewcommand\vec{\bm} 
\newcommand{\uGrad}{\vec{\nabla}}
\newcommand{\uRe}{{\mathrm{Re}}\,}
\newcommand{\uIm}{{\mathrm{Im}}\,}
\newcommand{\ud}{\,{\mathrm{d}}}

\newcommand{\uiiint}{\int\!\!\!\int\!\!\!\int}

\newcommand{\uvB}{\vec{B}} 
\newcommand{\uC}{C} 
\newcommand{\uD}{D} 
\newcommand{\ue}{e} 
\newcommand{\uE}{E} 
\newcommand{\ueE}{\epsilon} 

\newcommand{\ueEMS}{\epsilon_\mathrm{MS}} 
\newcommand{\ueIEMS}{\epsilon^\mathrm{I}_\mathrm{MS}} 
\newcommand{\ueIIEMS}{\epsilon^\mathrm{II}_\mathrm{MS}} 
\newcommand{\ugammaB}{\gamma_\mathrm{B}}
\newcommand{\uHopf}{{\cal{H}}} 
\newcommand{\uh}{h} 
\newcommand{\uH}{H} 
\newcommand{\uvH}{\vec{H}}     
\newcommand{\uvHD}{\vec{H}_\mathrm{D}} 
\newcommand{\uK}{K}         

\newcommand{\uvM}{\vec{M}}
\newcommand{\uvm}{\vec{m}}
\newcommand{\umi}{m_i}      
\newcommand{\umx}{m_{\mathrm{X}}}
\newcommand{\umy}{m_{\mathrm{Y}}}
\newcommand{\umz}{m_{\mathrm{Z}}}
\newcommand{\umr}{m_{\mathrm{\rho}}}
\newcommand{\umt}{m_{\mathrm{\theta}}}
\newcommand{\umf}{m_{\mathrm{\varphi}}}
\newcommand{\umIr}{m^\mathrm{I}_{\mathrm{\rho}}}
\newcommand{\umIt}{m^\mathrm{I}_{\mathrm{\theta}}}
\newcommand{\umIf}{m^\mathrm{I}_{\mathrm{\varphi}}}
\newcommand{\umIIr}{m^\mathrm{II}_{\mathrm{\rho}}}
\newcommand{\umIIt}{m^\mathrm{II}_{\mathrm{\theta}}}
\newcommand{\umIIf}{m^\mathrm{II}_{\mathrm{\varphi}}}

\newcommand{\uO}{O}         
\newcommand{\uP}{P_i}
\newcommand{\uPex}{P_\mathrm{EX}}
\newcommand{\uPa}{P_\mathrm{A}}
\newcommand{\uPz}{P_\mathrm{Z}}
\newcommand{\uPdm}{P_\mathrm{DM}}
\newcommand{\uPIdm}{P^\mathrm{I}_\mathrm{DM}}
\newcommand{\uPIIdm}{P^\mathrm{II}_\mathrm{DM}}
\newcommand{\uPms}{P_\mathrm{MS}}
\newcommand{\uPIms}{P^\mathrm{I}_\mathrm{MS}}
\newcommand{\uPIIms}{P^\mathrm{II}_\mathrm{MS}}
\newcommand{\up}{p_i}
\newcommand{\upex}{p_\mathrm{EX}}
\newcommand{\upa}{p_\mathrm{A}}
\newcommand{\upz}{p_\mathrm{Z}}
\newcommand{\updm}{p_\mathrm{DM}}
\newcommand{\upms}{p_\mathrm{MS}}

\newcommand{\uMs}{M_{\mathrm{S}}}
\newcommand{\uMssq}{M_{\mathrm{S}}^2}
\newcommand{\umuZ}{\mu_0}
\newcommand{\uq}{q} 
\newcommand{\uvs}{\vec{s}}  
\newcommand{\uX}{X}         
\newcommand{\uY}{Y}         
\newcommand{\uZ}{Z}         
\newcommand{\uXp}{\widetilde{X}} 
\newcommand{\uYp}{\widetilde{Y}} 
\newcommand{\uZp}{\widetilde{Z}} 
\newcommand{\uRp}{\widetilde{R}} 
\newcommand{\ur}{\rho}

\newcommand{\uvr}{\vec{\rho}}          
\newcommand{\uvrp}{\vec{\rho}'}  
\newcommand{\ut}{\theta}
\newcommand{\uf}{\varphi}

\begin{document}
\title{Magnetostatic bounds on stability of hopfions in bulk helimagnets}

\author{Konstantin L. Metlov}\email{metlov@donfti.ru}
\affiliation{Galkin Donetsk Institute for Physics and Engineering, R.~Luxembourg str.~72, Donetsk 283048, Russian Federation}

\date{\today}

\begin{abstract}
Magnetic hopfions are three-dimensional localized topological solitons in the volume of a magnet. In this work, starting with a classical free energy density of a helimagnet, an approximate variational model of hopfions is studied. The hopfion stability regions on the uniaxial anisotropy-external magnetic field phase diagram are computed and their evolution with increasing magnetostatic interaction strength is considered. 
It is found that magnetostatic interaction destabilizes the hopfions and, above the certain strength (relative to Dzyaloshinskii-Moriya interaction), destroys their stability completely. Numerical estimates for this bound are provided. They can help focus the search for materials, supporting bulk magnetic hopfions.
\end{abstract}


\maketitle

\section{Introduction}
\label{sec:intro}

Topology is a powerful tool for exploring the consequences of continuity. In our everyday observations of the physical world, continuity manifests itself as an evident gradual change of various physical quantities in space and time. Yet, as noticed in physics at least since lord Kelvin~\cite{thomson1869}, it also leads to emergence of a rich discrete set of objects. These correspond to the subdivision of a set of continuous mappings into topological classes, with their elements equivalent up to another (family of continuous maps) -- homotopy. The most ubiquitous example of these classes emerge during the classification of knots~\cite{knot_theory_book_1977} on a rope.

Rope is one-dimensional, but higher-dimensional knots are also possible. In magnetism these manifest themselves in the form of two-dimensional (2D) Belavin-Polyakov (BP) solitons~\cite{BP75} (the BP solitons in thin ferromagnetic films are known as magnetic bubbles~\cite{bobeck1975bubbles}, in bulk chiral magnets as skyrmions~\cite{RBP2006}, and in magnetic dots with in-plane magnetostatic anisotropy as magnetic vortices~\cite{UP93} even though they are described by a slightly different set of solutions of the same model, called merons~\cite{M10}), whose topological classification is the same as that of mappings between the surfaces of two spheres (denoted as $S^2\rightarrow S^2$). The first of these two spheres corresponds to all the locations in a 2D ferromagnet with periodic boundary conditions at infinity and the second to the endpoints of the constant-length local magnetization vectors. Just like magnetization configurations around Bloch points~\cite{doring68}, the BP solitons have a natural description in terms of functions of complex variable~\cite{BP75}, which becomes especially useful in restricted planar geometry~\cite{M10}.

A natural generalization of BP solitons to three dimensions~\cite{Faddeev1976} is to consider magnetization states, corresponding to the mapping of three-dimensional surface of a four-dimensional hypersphere $S^3$ (volume of the magnet with periodic boundary conditions at infinity) onto the surface of $S^2$ sphere (endpoints of the magnetization vectors). For a long time in mathematics it was accepted that all such mappings are homotopically equivalent (and topologically trivial) until Heinz Hopf in 1931 provided an example~\cite{Hopf1931} of a different (homotopically irreducible to the previously known) map. This example was generalized by Whitehead~\cite{whitehead1947}, who have shown that there is a number of such topologically non-trivial maps, characterized by an integer number -- Hopf invariant (or Hopf index). The topological solitons, corresponding to these non-trivial  $S^3\rightarrow S^2$ maps, took a name of hopfions. As far as magnetic hopfions are concerned, there was an extensive published theoretical research (including books) on those in bulk magnets with complex/unusual exchange interactions~\cite{Nicole1978,RKBDMB2022,blugel2023} and systems with uniaxial anisotropy~\cite{DI79,KIK90,BK11-book-v2}. There is recently a substantial renewed interest to magnetic hopfions~\cite{sutcliffe2018,lake2018}, which produced several experimental realizations of magnetic hopfions (or parts of hopfions) in confined geometry~\cite{kent2021,altbir2023,zheng2023}.

Yet, only the hopfions in bulk magnets can fully realize unique potential of these topological objects for 3D spintronics and search for them still continues. These bulk hopfions are not easy to simulate numerically, because the amount of computational resources, necessary to fully simulate 3D magnets, scales much faster with system size (which must be reasonably high to simulate the bulk), compared to the extensively studied in simulations planar nanomagnets. The hopfions do appear in micromagnetic simulations (e.g.~\cite{lake2018}), but analytical (or semianalytical) models are still better suited for mapping the hopfion stability range on the magnetic phase diagram.

Such a semianalytical model (but still based on classical micromagnetic Hamiltonian) was considered recently in~\cite{M2023_TwoTypes}. It predicts the existence of two types of metastable magnetic hopfions in a bulk helimagnet. The present work reports a comprehensive study of the dependence of the hopfion stability region (for both types of hopfions) on the relative strength of the magnetostatic interaction. In the next section, the considered model is formally introduced, the corresponding magnetic energy terms are computed and presented in Appendices.
In Section~\ref{sec:results}, numerically computed stability ranges of the hopfions are plotted and discussed. Section~\ref{sec:conclusions} concludes the study.

\section{Model}
\label{sec:model}
As a starting point, consider the ansatz for an $S^3\rightarrow S^2$ mapping with non-zero Hopf index
\begin{equation}
\label{eq:Whitehead}
w=e^{\imath\chi}\frac{u^n}{v^k},
\end{equation}
where $u$ and $v$ are complex coordinates on a unit $S^3$ sphere, satisfying the constraints $|u|^2+|v|^2=1$; $n$ and $k$ are two mutually prime integers; $\imath=\sqrt{-1}$ and $w$ is the coordinate on the target $S^2$ sphere, so that the components of the normalized magnetization vector $\uvm=\uvM/\uMs$ are
\begin{equation}
\label{eq:stereo}
\{\umx+\imath\umy,\umz\}=\{2w, 1 - |w|^2\}/(1+|w|^2). 
\end{equation}
The Hopf index of the resulting map is $\uHopf=n k$, but we will limit ourselves here only to the simplest $n=k=1$ case, whose Hopf index of $\uHopf=1$ is directly confirmed in the  \ref{sec:HopfIndex}. The real parameter $\chi$ does not influence the topology and~\eqref{eq:Whitehead} coincides with the Whitehead's ansatz~\cite{whitehead1947} when $\chi=0$. Yet, the energy analysis~\cite{M2023_TwoTypes}, based on a classical micromagnetic Hamiltonian (including most notably the Dzyaloshinskii-Moriya and the magnetostatic energy terms), shows that $\chi$ takes one of two stable equilibrium values: $\chi=\pi/2$ and $\chi=3\pi/2$. These values correspond to type~I and type~II hopfions, respectively.

\begin{figure}[tb]
\begin{center}
\includegraphics[width=\columnwidth]{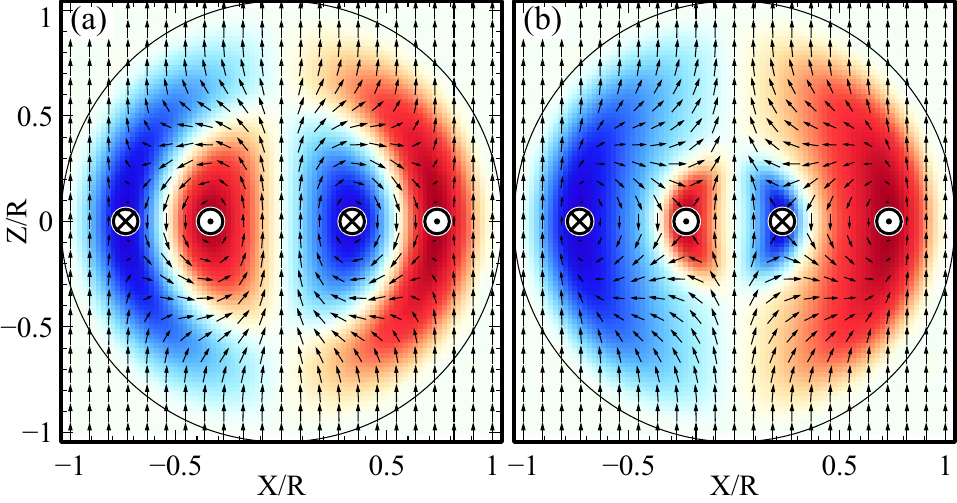}
\end{center}
 \caption{\label{fig:hopfillustr} Magnetization distributions inside the equilibrium type~I (a) and type~II (b) hopfions when no field or anisotropy is present ($h=q=0$) and the magnetostatic parameter $\mu=1/3$; the shading and circular symbols show the the out-of-plane magnetization component; the equilibrium size parameters in this case are $\nu^\mathrm{I}\approx\num{-0.199894}$ and $\nu^\mathrm{II}\approx\num{0.099267}$, so that the type~II hopfion is roughly twice the size of the type~I hopfion.}
\end{figure}

The coordinates on the sphere $S^3$ can be specified in terms of the coordinates of a point $\uXp$, $\uYp$, $\uZp$ on the extended Euclidean space $E^3$
\begin{equation}
\label{eq:R3S3}
u=\frac{2(\uXp+\imath \uYp)R}{\uXp^2+\uYp^2+\uZp^2+R^2}, \quad
v=\frac{R^2-\uXp^2-\uYp^2-\uZp^2+\imath 2 \uZp R}{\uXp^2+\uYp^2+\uZp^2+R^2},
\end{equation}
where $R$ is a (yet unknown) spatial scale. 
The space $E^3$ implies periodic boundary conditions at infinity (so that going to the infinity in any direction leads to the same single ``infinity'' point). The mapping~\eqref{eq:R3S3} is just a stereographic projection, which maps $E^3$ to $S^3$, similar to the stereographic projection, which maps the extended complex plane $E^2$ to the Riemann's sphere $S^2$. One can see that the origin $\uXp=\uYp=\uZp=0$ is mapped onto $(u,v)=(0,1)$ and the infinitely distant point in any direction $\uXp,\uYp,\uZp\rightarrow\infty$ to $(u,v)=(0,-1)$. Both these points map to $w=0$ and $\uvm=\{0,0,1\}$ directed along the $\uO\uZ$ axis of the chosen Cartesian coordinate system. Note that the a single hopfion, specified by the Whitehead's ansatz~\eqref{eq:Whitehead}, occupies the whole $S^3$ or the whole $E^3$ space. On the other hand, one of the important characteristics of the topological soliton -- its strong localization in space. This can be achieved by mapping the physical $\uX$, $\uY$, $\uZ$ space onto $E^3$ using
\begin{equation}
 \label{eq:StoR}
 \{\uXp,\uYp,\uZp\}=\frac{\{\uX,\uY,\uZ\}}{1-f(\sqrt{\uX^2+\uY^2+\uZ^2}/R)},
\end{equation}
where $f(\ur)$, $0\le \ur\le 1$ is a (yet unknown) function, satisfying the boundary conditions $f(0)=0$, $f'(0)=0$, $f(1)=1$. This ensures that the hopfion is fully contained inside the ball of the radius $R$, whose boundary $\uX^2+\uY^2+\uZ^2=R^2$ is mapped to the infinitely distant point $\uXp,\uYp,\uZp\rightarrow\infty$ in $E^3$. In the framework of this Ritz-type model, finding stable hopfions means finding the value of the scalar parameter $R$ and the profile function $f(\ur)$, minimizing the Hamiltonian of the physical model, expressed in terms of the vector $\uvm$ of the fixed length $|\uvm|=1$.
Explicitly, the magnetization components inside the $\uHopf=1$ hopfion are given in the \ref{sec:mspher} and shown in Fig.~\ref{fig:hopfillustr}.

Note that only the last equation~\eqref{eq:StoR} carries the model assumptions here. This is because all the hopfions with the same $\uHopf$ are equivalent up to a continuous map (homotopy). A sufficiently general map from $\uX$, $\uY$, $\uZ$ to $\uXp$, $\uYp$, $\uZp$ together with~\cref{eq:Whitehead,eq:stereo,eq:R3S3} is able to describe the exact hopfion solutions of a considered physical model Hamiltonian (not even necessarily micromagnetic). Limiting the minimization to a specific subset of the maps~\eqref{eq:StoR} simplifies the consideration, but makes the resulting hopfion profiles only approximate.

The map~\eqref{eq:StoR} is sufficient~\cite{M2023_TwoTypes} to obtain stable hopfion solutions for the classical micromagnetic free energy density
\begin{equation}
\label{eq:energy} 
\begin{aligned}
        \ue = & \frac{\uC}{2}\sum_{i=\uX,\uY,\uZ} \left|\uGrad\umi\right|^2 +\uD\, \uvm\cdot[\uGrad\times\uvm] -\\
        &   -\umuZ \uMs \left(\uvm \cdot \uvH\right) - \frac{\umuZ\uMs}{2}\left(\uvm \cdot \uvHD\right)- \uK\left(\uvm \cdot \uvs\right)^2, 
\end{aligned}
\end{equation}
where $C=2A$ is the exchange stiffness, $D$ is the Dzyaloshinskii-Moriya interaction strength~\cite{BS1970,BJ1980}, $\uvH$ is the external magnetic field, $\uvHD$ is the demagnetizing field, created by all the magnetic moments in the material as per Maxwell's equations, $\uK$ and $\uvs$ are the uniaxial anisotropy constant and director. Following Aharoni~\cite{AharoniBook}, the relation  $\uvB = \umuZ(\uvH + \ugammaB \uvM)$ for the magnetic induction allows us to cover all the common systems of magnetic units ($\umuZ$ is the permeability of vacuum and  $\ugammaB=1$ in SI; $\ugammaB=4\pi$, $\umuZ=1$ in CGS). In the present consideration we will assume that the direction of the external field $\uvH=\{0,0,\uH\}$ and the anisotropy axis director $\uvs=\{0,0,1\}$ coincide with the hopfion axis lying along the $\uO\uZ$ axis of the Cartesian coordinate system. In this system, according to \cref{eq:Whitehead,eq:stereo,eq:R3S3,eq:StoR}, the magnetization along the axis of the hopfion $\uX=\uY=0$ on its spherical boundary and outside is $\uvm=\{0,0,1\}$.

Assuming (as in~\cite{M2023_TwoTypes}) the hopfions are arranged in a close-packed lattice, its total energy $\uE=(1/V)\uiiint_V e \ud^3\vec{r}$ per unit cell volume $V$ in dimensionless units $\ueE=\uE \uC/\uD^2$ can be expressed as a functional $\ueE = \int_0^1 F \ud\ur$ with
\begin{equation}
\label{eq:energyPerCell}
\begin{aligned}
 F = \frac{1}{4\sqrt{2}} \biggl(& \nu^2\uPex+\nu\uPdm+\frac{q}{2}(\uPa-4\sqrt{2})+\\
 &+h(\uPz-4\sqrt{2})+\mu^2\uPms \biggr),
\end{aligned}
\end{equation}
where $4\sqrt{2}R^3$ is the volume per hopfion inside the unit cell of the close packed lattice, $\nu=\uC/(\uD R)$ is the dimensionless size parameter, inversely proportional to the hopfion radius $R$, $\uh=\umuZ\uMs\uC\uH/\uD^2$ is the normalized external field, $\uq=2\uC\uK/\uD^2$ normalized anisotropy quality factor, $\mu^2=\umuZ\ugammaB\uMssq\uC/\uD^2$ is the dimensionless magnetostatic interaction strength parameter ($\mu^2$ is also known as the susceptibility of the conical helix~\cite{GWG2017}) and the energy functions
\begin{equation}
\label{eq:uP}
\uP = \int\limits_0^{2\pi}\ud \varphi\int\limits_0^{\pi}\ud \theta\, \up\, \ur^2\sin\theta
\end{equation}
with $\upex=(R^2/2) \sum_{i=\uX,\uY,\uZ}|\uGrad\umi|^2$, $\updm=R\, \uvm\cdot[\uGrad\times\uvm]$, $\upa=1-\umz^2$, $\upz=1-\umz$, while $\upms$ is itself defined via the integral over $\ur$ due to non-locality of the dipolar energy. For both types of hopfions, the angular integrals in~\eqref{eq:uP} can be evaluated analytically. The expressions for $\uP$ are given in the \ref{sec:energyFuncs}. They depend on $\ur$, the hopfion profile function $f(\ur)$ and its derivative $f'(\ur)$.

The problem for finding the equilibrium $f(\ur)$ then reduces to solving the Euler-Lagrange equation
\begin{equation}
\label{eq:EulerLagrange}
 \frac{\partial F}{\partial f} - \frac{\partial}{\partial\ur}\frac{\partial F}{\partial f'} = 0.
\end{equation}
In our case, it is a second order differential equation, requiring two boundary/initial conditions to solve. There is also an unknown scalar parameter $\nu$, which needs to be chosen in such a way that all three initial/boundary conditions
\begin{equation}
 \label{eq:boundary}
 f(0)=0,\quad f'(0)=0,\quad f(1)=1
\end{equation}
are satisfied. This can be implemented by making $\nu$ a function $\nu(\rho)$ and adding another equation:
\begin{equation}
 \label{eq:nuEqu}
 \nu'(\rho)=0.
\end{equation}
Together with~\eqref{eq:EulerLagrange} we get a system of three first order differential equations for $f(\rho)$, $f'(\rho)$ and $\nu(\rho)$. Three initial/boundary conditions~\eqref{eq:boundary} make this problem well defined. It can be solved numerically using the method of shooting. Here this was done with MUS~\cite{MUS1984}.

Non-locality of magnetostatic interaction poses another problem by adding an integral term to the Euler equation~\eqref{eq:EulerLagrange}. It was handled using an iterative scheme. When solving the boundary value problem in \cref{eq:EulerLagrange,eq:boundary,eq:nuEqu} for $f_n(\ur)$ on the $n$-th iteration, the magnetostatic functions $W_1$, $V_1$, $V_2$ (see the \ref{sec:energyFuncs}) are evaluated, based on the previous iteration $f_{n-1}(\ur)$. This removes the integral term from the equation, replacing it by an integral of the known previous-iteration function. The calculation is then repeated until the changes between the hopfion profile on successive iterations become negligible. Once the convergence is achieved, the resulting hopfion profile solves the full integro-differential Euler-Lagrange equation.

\section{Results and discussion}
\label{sec:results}
There are two instabilities, embedded into the considered model: hopfions may be unstable with respect to the collapse of their size to zero $\nu\rightarrow\infty$ or to the expansion to infinity $\nu\rightarrow0$. In either case, the method of shooting fails to find the solution for $f(\ur)$, satisfying the boundary conditions~\eqref{eq:boundary}. If the magnetostatic interaction is included, near the stability boundary the iterations for finding the $f(\ur)$, satisfying the integro-differential Euler equation, may not converge to the true solution and instead loop around it (in such situation we also consider the solution to be unstable). 

The process of tracing the stability boundary starts at a certain point of the $q$--$h$ phase diagram for a particular value of $\mu$, such that the hopfion solution at this point exists and is stable. Next step is to move in an arbitrary direction in small and gradually decreasing steps, using the previous $f(\ur)$ and $\nu$ as initial guess for the solution in the next point, until the solution becomes unstable. This means the stability boundary is reached. Next, the tracer follows the boundary, keeping it always on a right (or left) side of its current position until a full circle is made. The result is a complete stability line on the phase diagram. This procedure (together with computation of the equilibrium hopfion profile, described in the previous section) is implemented in Fortran programming language and published as a GitHub repository under the project name {\tt magnhopf}~\cite{magnhopf}.

\begin{figure}[tb]
\begin{center}
\includegraphics[width=0.6\columnwidth]{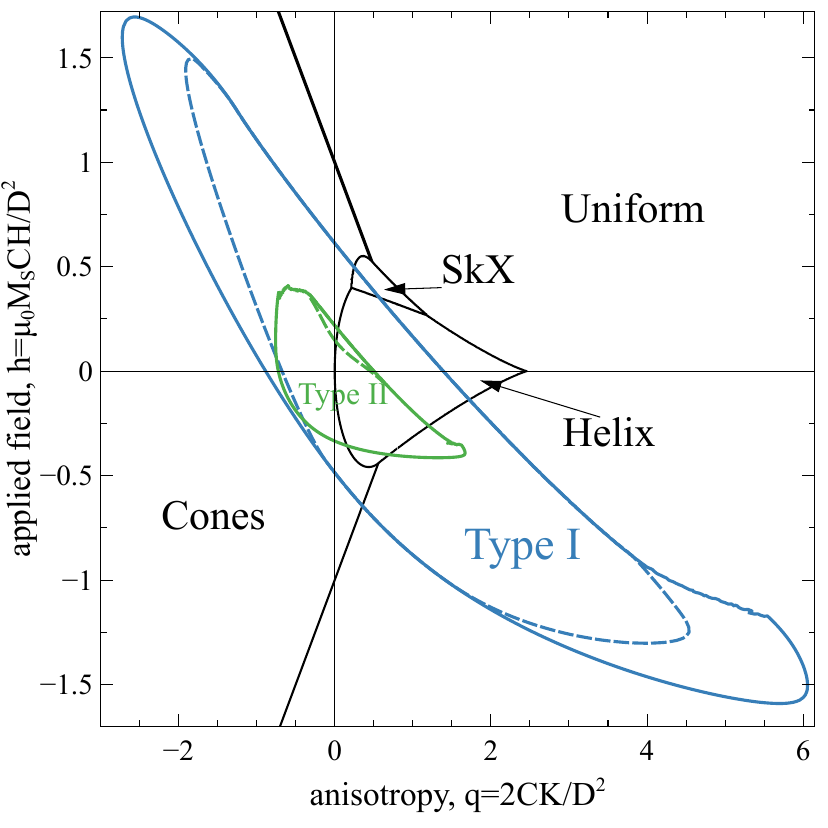}
\end{center}
 \caption{\label{fig:stabmuzero}Theoretical stability regions of the type~I and type~II hopfions when the magnetostatic interaction is neglected ($\mu^2=0$), superimposed over the ground state diagram of the helimagnet with uniform, conical, helical, and skyrmion lattice phase~\cite{BY1989,BH1994}. The dashed lines separate the hopfion states with monotonous and non-monotonous $f(\ur)$.}
\end{figure}

The resulting stability regions on the $q$--$h$ plane when the magnetostatic interaction is negligible ($\mu=0$) are shown in Fig.~\ref{fig:stabmuzero}.
Due to the approximate nature of the performed search, there is no guarantee that the true stability boundary is found this way. The strong statement is that anywhere inside this stability boundary the above-described numerical procedure can always obtain a stable hopfion solution with a particular hopfion profile $f(\ur)$ and a hopfion size parameter $\nu$ minimizing the total magnetic energy~\eqref{eq:energyPerCell} as per~\eqref{eq:EulerLagrange}. Improvement in the numerical procedure, such as increasing the search resolution, might move the stability lines somewhat. Nevertheless, the numerical experiments (with different resolutions and other numerical parameters) do not show that this dependence is significant and the resulting fluctuations on the stability lines are small.

Compared to Fig.~2 in~\cite{M2023_TwoTypes}, this diagram covers also negative values of the quality factor and the external field, fully outlining the stability regions of hopfions. One can see from the superimposed classical ground state diagram of the helimagnet that (meta) stable hopfions exist roughly at the same range of material parameters as the skyrmions and other well known phases.

\begin{figure}[tb]
\begin{center}
\includegraphics[width=\columnwidth]{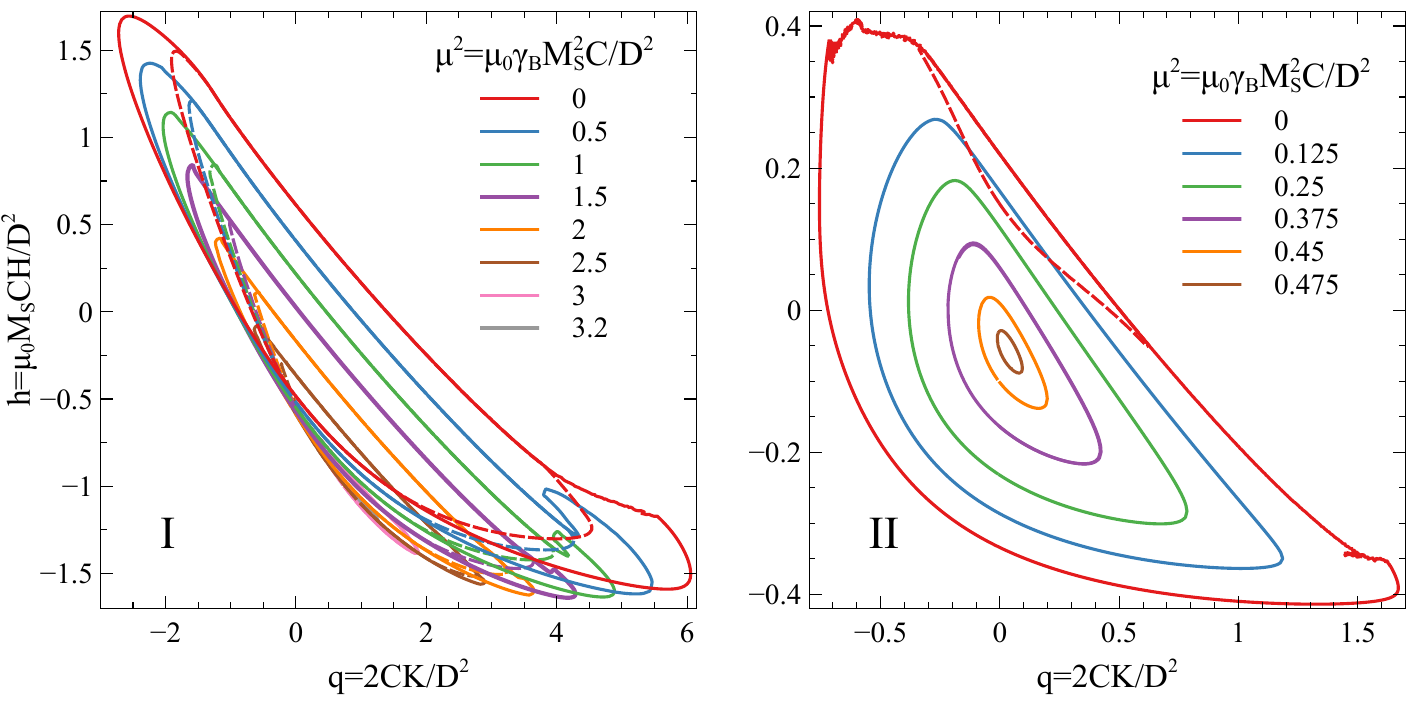}
\end{center}
 \caption{\label{fig:stabmu}Evolution of stability regions of the type~I (left) and type~II (right) hopfions as the magnetostatic interaction strength parameter $\mu^2$ increases.}
\end{figure}

In~\cite{M2023_TwoTypes} it was already noted that the equilibrium hopfion profile $f(\ur)$ is not always monotonous. This is not a problem, since a hopfion with a non-monotonous $f(\ur)$ can be transformed into one with monotonous profile via a homotopy, meaning that non-monotonicity does not change the topological class. Nevertheless, in Fig.~\ref{fig:stabmuzero} and Fig.~\ref{fig:stabmu} the subset of hopfion states with monotonous profiles is additionally outlined with the dashed line. This subset for both hopfion types is located on top (larger field side) of each stability region and for the type~II hopfions at $\mu^2=0$ it is very small. For larger $\mu^2$ a subset of type~II hopfions with monotonous $f(\ur)$ quickly vanishes, it was impossible to find already for $\mu^2\ge0.125$ in the present numerical calculations.

For some time, since the appearance of the famous Hobart-Derrick theorem~\cite{Hobart1963,*Derrick1964}, it was believed that 3D solitons in the bulk are unstable and their equilibrium size is not defined. Until Faddeev pointed out the importance of chirality for the soliton stabilization~\cite{Faddeev1976}, which breaks the applicability of the Hobart-Derrick theorem. The chiral Dzyaloshinskii-Moriya interaction, considered in the present work, is also sufficient to stabilize finite size hopfions in the bulk. Or, for the total energy~\eqref{eq:energyPerCell} to have a minimum at a particular value of the dimensionless hopfion size parameter $\nu$.

On the other hand, unlike the well known ground states of the helimagnet (helices, cones and skyrmions), hopfions, consisting of a pair of {\em curved} vortex and antivortex lines, do produce a non-negligible amount of magnetic charges (whose volume density is proportional to the $\uGrad\cdot\uvm$). Because the energy of these charges is strictly positive and its contribution in the total energy~\eqref{eq:energyPerCell} is scaled by the factor $\mu^2$, it is natural to expect that increase in $\mu$ would eventually destabilize the hopfions, forcing the magnetization to assume another divergence-free lower energy configuration. This is exactly what our numerical results show for the dependence of the hopfion stability regions on the magnetostatic interaction strength $\mu^2$, which is plotted in Fig.~\ref{fig:stabmu}.
The main feature on these plots is that the stability region progressively shrinks (in the case of type~I hopfions it also moves towards the smaller/negative fields and larger anisotropy) as the magnetostatic interaction strength increases. It means that the magnetostatic interaction suppresses the hopfion state.

This is interesting to consider in context of a known theoretical result~\cite{M01_CT,CMSGHBRHCG2020} that the magnetostatic interaction can stabilize the cross-tie domain wall in the limit of infinite film thickness (when it is just a linear sequence of alternating parallel infinite straight vortex and antivortex lines in the bulk of the magnet). Since the $\uHopf=1$ hopfion is just a combination of a vortex and an anti-vortex lines, woven on top of each other, the deciding factor seems to be the curvature of these lines. The straight nearby vortex and antivortex lines attract and form a bound state (the cross-tie wall) via the attraction of the volume magnetic charges of the opposite sign around them. But the circular vortex and antivortex lines in the hopfion produce more of the concentrated magnetic charges of the same sign, increasing magnetostatic energy up to a point that it becomes energetically favorable to remove the hopfion altogether. This is like in the case of the ordinary magnetic domain walls, which have associated area energy density. This energy disfavors the wall curvature and produces additional curvature-related pressure from the wall (like e.g. in the case of a circular domain wall of a magnetic bubble, where this pressure eventually leads to the bubble collapse if its radius decreases below a certain threshold).

Whatever the mechanism, from the evolution of the stability regions in Fig.~\ref{fig:stabmu} it is possible to roughly estimate the limiting values of $3.2<\mu_\mathrm{I}^2\lessapprox3.3$ and $0.475<\mu_\mathrm{II}^2\lessapprox0.48$. At $\mu>\mu_\mathrm{I}$ (for type~I hopfions) and $\mu>\mu_\mathrm{II}$ (for type~II hopfions) the stability region completely vanishes and the hopfion state becomes impossible to stabilize in our computations. These numbers are in range of the parameters of well known chiral skyrmion hosting materials, shown in Table~\ref{tab:matparam}.
\begin{table}
\begin{center}
\begin{tabular}{lrr}
 \hline\hline
 Material & $\mu^2$ & $T_\mathrm{C}$, K \\
 \hline
 $\mathrm{Mn}\mathrm{Si}$ & 0.34 & 29 \\
 $\mathrm{Cu}_2\mathrm{O}\mathrm{Se}\mathrm{O}_3$ & 1.76 & 58 \\
 $\mathrm{Fe}_{0.8}\mathrm{Co}_{0.2}\mathrm{Si}$ & 0.65 & 28 \\
 $\mathrm{Fe}\mathrm{Ge}$ & 3.43 & 278 \\
\end{tabular}
\end{center}
 \caption{\label{tab:matparam}Material parameters for some skyrmion hosting materials, collected in~\cite{GWG2017} from references therein.}
\end{table}
It follows, for example, that the room temperature FeGe is unlikely to support the free standing bulk hopfions of any type. But other listed materials can support either type~I hopfions or both (like MsSi) at cryogenic temperatures. Let us stress, that this limit concerns only the free standing hopfions in the bulk. Stabilization of pinned hopfions in nanostructures~\cite{lake2018} or observations of individual pars of hopfions (like torons~\cite{altbir2023}, which are just a single circular vortex line) are not subject to this particular limit.

Stability of magnetic hopfions when the magnetostatic interaction is neglected is also very well established via the numerical simulations~\cite{sutcliffe2018,TS2018}. It is important to note, however, that, finite-element (or finite-difference) micromagnetic approach is by design best suited for consideration of systems with restricted geometry. While this works extremely well for simulating planar magnetic nanostructures, applying it to a 3D problem entails steep computational cost. That's why a full numerical simulation of the 3D lattice of hopfions, like the close-packed one, considered in the present work, to author's knowledge was not achieved yet. There is a simulation of a stable two-dimensional close-packed lattice of hopfions in a thin film~\cite{TS2018}, confined by the additional surface anisotropy (which mimics the effect of other layers of hopfions in a 3D lattice). It shows that, despite the additional pressure, imposed by confinement, the hopfions keep their stability. To abolish the confinement and make a full 3D simulation of hopfions it is necessary to have a good initial configuration for the micromagnetic simulation and an equally good initial estimate for all three periods of the hopfion lattice. The present semianalytical model and its further development can lead to such estimates, but even then tracing the whole phase diagram with magnetostatic energy included and comparable resolution will, probably, not be possible with finite element/difference simulations in the near future.

Another recent numerical result, which can be considered a confirmation of the present calculation, is the observation that in the simulation of the hopfion rings~\cite{blugel2023} around a vortex tube, the rings lose their stability, once the ends of the tube are detached from the nanocylinder's surfaces. Thus, once the hopfions become fully localized in this numerical experiment, they immediately lose their stability. This is easy to understand on the basis of the present work, because the simulation in~\cite{blugel2023} included the magnetostatic energy and was done for the FeGe, whose dimensionless parameter $\mu$ at room temperature is outside of the hopfion stability range (see Table~\ref{tab:matparam}). Even if a lower value of $\mu$ is assumed due to the temperature in the simulations~\cite{blugel2023} being lower than the room temperature, the Type~I hopfion stability range (see Fig.~\ref{fig:stabmu}) quickly shrinks above $\mu\gtrsim2$ and for $\mu\gtrsim3$ the hopfions are always unstable in the case of zero anisotropy $q=0$, considered in~\cite{blugel2023}.

Let us also note that the above estimates for $\mu_\mathrm{I}$ and $\mu_\mathrm{II}$ depend on the considered variational model and can be improved by adding more degrees of freedom. In particular, the hopfion's axial symmetry suggests the possibility of elliptical (in any of the cross-sections, passing through the hopfion's symmetry axis) deformation. Such analysis was performed by adding a scalar aspect ratio free parameter to the present variational model at $\mu=0$ (i.e. neglecting the magnetostatic energy). It shows that, while the equilibrium hopfions do deform elliptically, they find an energy minimum at a particular finite non-zero aspect ratio and remain stable. Such an additional degree of freedom decreases the energy of $\mu=0$ hopfions and  slightly extends their stability range on the anisotropy-field phase diagram. These preliminary findings were reported at \cite{M2024Kourovka} and will be presented in a forthcoming publication. Still, no other estimates of $\mu_\mathrm{I}$ and $\mu_\mathrm{II}$ are currently known and the present estimates, based on the spherical hopfion model, already provide a substantial practical value for the potential hopfion-hosting material selection.

Thus, the magnetostatic interaction has a destabilizing influence on the hopfions (unlike cones, helices or skyrmions, which have negligible magnetostatic energy) and makes the observation (in nature or numerical experiments) of free standing hopfions harder. So, the ideal material to look for hopfions should have as small $\mu^2=\umuZ\ugammaB\uMssq\uC/\uD^2$ as possible. Probably the most straightforward way to achieve this is via a competition of the sublattice magnetizations in a chiral ferrimagnet or may be even in a chiral altermagnet~\cite{SSJ2022}.

\section{Summary \& Conclusions}
\label{sec:conclusions}
A variational model for magnetic hopfions in a classical helimagnet is considered. Its energy terms, corresponding to the exchange, uniaxial anisotropy, external field, Dzyaloshinskii-Moriya (DM) interaction and the magnetostatic energy are computed and presented in the \ref{sec:energyFuncs}. A program for finding the equilibrium hopfion profile, minimizing the corresponding energy functional, is developed~\cite{magnhopf}. This program allows to compute the full hopfion stability region on the anisotropy-external field phase diagram and study its dependence on the magnetostatic interaction strength to the DM interaction constant ratio $\mu^2$. It is shown that the magnetostatic interaction has destabilizing effect on hopfions. Moreover, there are certain limiting values of $\mu^2$, specific to each of the two hopfion types, such that for higher $\mu^2$ the hopfion stability region on the phase diagram completely vanishes. These limiting values are estimated on the basis of the studied model and are discussed in context of the material parameters for certain well known model helimagnets. They can be useful to help narrow the search for stable bulk hopfions. 

\appendix
\section{Hopf index and fibrations}
\label{sec:HopfIndex}

One way to confirm the Hopf index of the magnetization distributions, specified by \cref{eq:Whitehead,eq:stereo,eq:R3S3,eq:StoR}, is to use the Whitehead's expression~\cite{whitehead1947} for it. However, unlike a similar integral expression for the 2D topological charge, the Whitehead's integral involves a vector potential for the emergent magnetic field, which, moreover, needs to be specified in a particular gauge. There is a simpler and more direct approach, which is based on the analysis of pre-images of the $S^3\rightarrow S^2$ map, defined by the magnetization distribution in the bulk.

The $S^3$ sphere represents a set of spatial coordinates in the bulk with periodic boundary conditions at infinity, while the $S^2$ sphere contains the endpoints of the fixed length magnetization vectors. The source $S^3$ sphere has one extra dimension, compared to the target $S^2$ sphere. It means that (for the smooth maps) pre-image of every point on the target $S^2$ sphere (the set of points whose local magnetization points towards exactly the same direction) is a curve (a one-dimensional object). The whole set of these curves (fibers) for all the points on the target $S^2$ sphere is called ``fibration''.  The defining property of the Hopf map~\cite{Hopf1931} is that the pre-image curves for any two different magnetization directions are interlinked exactly $\uHopf$ times.

For the considered analytical model the pre-images can be easily computed. Specifically, one can verify via the direct substitution that the curve on the $S^3$ sphere
\begin{equation}
 \label{eq:preimagesS3}
 u=u_0 e^{\imath k \alpha}, \quad
 v=v_0 e^{\imath n \alpha}, \quad
\end{equation}
where $u_0$ and $v_0$, $|u_0|^2+|v_0|^2=1$ are complex constants, is the solution of~\eqref{eq:Whitehead} for any $0<\alpha<2\pi$.  It parametrically defines a closed curve on $S^3$, whose every point is mapped to a particular constant direction of the magnetization $w_0=e^{\imath\chi}u_0^n/v_0^k$. In the $E^3$ space, defined by the map~\eqref{eq:R3S3} these curves can be expressed as
\begin{equation}
 \uXp+\imath \uYp =\frac{R\,u_0 e^{\imath k \alpha}}{1+\uRe(v_0 e^{\imath n \alpha})},
 \uZp =\frac{R\,\uIm(e^{\imath n \alpha}v_0)}{1+\uRe(e^{\imath n \alpha}v_0)}.
\end{equation}
It is possible to eliminate $u_0$ and $v_0$ by considering a fiber, which passes through a specified point $\{\uXp_0, \uYp_0, \uZp_0 \}$ at $\alpha=0$ and connects it to all the other points with the same direction of the magnetization. Such a fiber is given by
\begin{equation}
 \begin{aligned}
  \uXp & = \frac{2 R^2 \left(\uXp_0 \cos (\alpha  k)-\uYp_0 \sin (\alpha  k)\right)}{\left(R^2-\uRp_0^2\right) \cos (\alpha n)-2 R \uZp_0 \sin (\alpha  n)+R^2+\uRp_0^2}, \\
  \uYp & = \frac{2 R^2 \left(\uXp_0 \sin (\alpha  k)+\uYp_0 \cos (\alpha  k)\right)}{\left(R^2-\uRp_0^2\right) \cos (\alpha n)-2 R \uZp_0 \sin (\alpha  n)+R^2+\uRp_0^2}, \\
  \uZp & = \frac{R \left(\left(R^2-\uRp_0^2\right) \sin (\alpha  n)+2 R \uZp_0 \cos (\alpha n)\right)}{\left(R^2-\uRp_0^2\right) \cos (\alpha  n)-2 R \uZp_0 \sin (\alpha  n)+R^2+\uRp_0^2},
 \end{aligned}
\end{equation}
where $\uRp_0^2=\uXp_0^2+\uYp_0^2+\uZp_0^2$. It can be transformed to physical space using the map~\eqref{eq:StoR}, which includes the hopfion profile function $f(\ur)$.

\begin{figure}[tb]
\begin{center}
\includegraphics[width=\columnwidth]{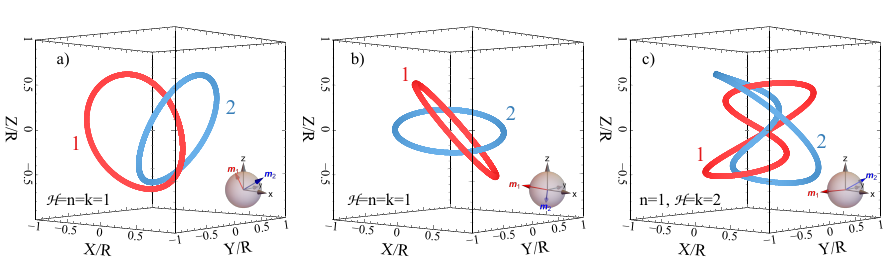}
\end{center}
 \caption{\label{fig:fibers}Pre-images~\eqref{eq:preimagesphys} for $\alpha\in[0,2\pi]$ and the following pairs of points $\{\ur_0,\uf_0,\ut_0\}$: a) $\{1/2,0,2\pi/6\}$ and $\{1/4,-\pi/2,\pi/2\}$; b) $\{1/2,-\pi,\pi/4\}$ and $\{1/2,\pi/4,\pi/2\}$; c) $\{1/2,0,3\pi/4\}$ and $\{1/4,-\pi/8,\pi/2\}$; the corresponding two magnetization vector directions (assuming $\chi=\pi/2$) are schematically shown in the lower right corner of each subfigure.}
\end{figure}

For the purpose of illustration, let us use a simple analytical profile $f(\ur)=\ur^2$, which satisfies all the necessary boundary condition. This profile is similar to the equilibrium Type~I hopfion at zero field and anisotropy. Using more realistic equilibrium profiles will deform the pre-images in physical space, compared to the present illustrative example, but their interlinking will remain the same. Another advantage of considering a simple quadratic profile is that the corresponding map~\eqref{eq:StoR} can be easily analytically inverted. This allows to compute the pre-image $\uvr=\{\uX,\uY,\uZ\}/R$, passing through a point $\{\ur_0,\uf_0,\ut_0\}$, specified by its spherical coordinates in physical space, explicitly
\begin{equation}
\label{eq:preimagesphys}
 \begin{aligned}
 \uvr & =
\frac{R \left(1-\sqrt{\frac{8 c}{b+c}-3}\right)}{b-c} \biggl(\rho _0 \left(1-\rho _0^2\right)
 \bigl\{ \sin \left(\theta _0\right) \cos \left(\alpha  k+\phi _0\right), \\
 &\sin \left(\theta _0\right) \sin \left(\alpha  k+\phi _0\right), \cos \left(\theta _0\right) \cos (\alpha 
   n)\bigr\} + \bigl\{0,0,\frac{\left(c-2
   \rho _0^2\right) \sin (\alpha  n)}{2}
   \bigr\}\biggr), \\
   b&=2 \rho _0 \left(\rho _0^2-1\right) \cos \left(\theta _0\right) \sin (\alpha  n)+\left(\rho _0^4-3 \rho
   _0^2+1\right) \cos (\alpha  n), \\
   c&=\rho _0^4-\rho _0^2+1.
 \end{aligned}
\end{equation}
Some of these fibers are plotted in Fig.~\ref{fig:fibers}.
The two sub-figures from the left show two particular pairs of preimages for the $\uHopf=1$ case, which is considered in the present paper. The right sub-figure shows the case of $\uHopf=2$ and the two shown pre-images wind around each other exactly two times. Note that to have a non-trivial Hopf index it is important that {\em all} of the pre-images for any distinct pair of magnetization directions (or the starting points) are interlinked exactly $\uHopf$ number of times. This is the case for the magnetization distributions from \cref{eq:Whitehead,eq:stereo,eq:R3S3,eq:StoR} for any pair of different $\{\ur_0,\uf_0,\ut_0\}$ in~\eqref{eq:preimagesphys}, not just for the few examples shown. The whole set of pre-images is exactly the same for Type~I and Type~II hopfions, but the actual directions of the magnetization vectors along pre-images of different hopfion types, passing through the same initial point, differ by a uniform rotation.

\section{Hopfion magnetization in spherical coordinates}
\label{sec:mspher}
Because of the symmetry, a spherical coordinate system with a polar axis along the direction of the applied field (and the considered anisotropy) is natural for description of spherical hopfions. 
Starting from the connection between the Cartesian and spherical coordinates
\begin{equation}
\label{eq:spherical}
\{\uX,\uY,\uZ\} = \ur R \{ \cos\uf \sin\ut,
\sin\uf \sin\ut, \cos\ut\},
\end{equation}
the transformation matrix, connecting the Cartesian and spherical vectorial components, consists of normalized (to have the unit length) rows of derivatives of the coordinate transforms~\eqref{eq:spherical} over $\ur$, $\ut$, $\uf$
\begin{equation}
 \label{eq:sphericalvec}
 \begin{bmatrix}
  \umr \\ \umt \\ \umf
 \end{bmatrix}
 =
 \begin{bmatrix}
  \sin\ut \cos\uf & \sin\ut \sin\uf & \cos\ut \\
  \cos\ut \cos\uf & \cos\ut \sin\uf & -\sin\ut\\
 -\sin\uf & \cos \uf & 0 \\
 \end{bmatrix}
 \begin{bmatrix}
  \umx \\ \umy \\ \umz
 \end{bmatrix}.
\end{equation}

The Cartesian magnetization vector components $\{\umx,\umy,\umz\}$ can be computed from \cref{eq:Whitehead,eq:stereo,eq:R3S3,eq:StoR}. The same vector field in the spherical coordinate system~\eqref{eq:spherical}, \eqref{eq:sphericalvec} for hopfions of the first type ($\chi=\pi/2$) is
\begin{equation}
 \{\umIr,\umIt,\umIf\} =
 \left\{\cos\ut, -\frac{\left(g^4-6 g^2 \ur ^2+\ur ^4\right) \sin\theta}{\left(g^2+\ur ^2\right)^2}, \frac{4 g \ur  \left(g^2-\ur ^2\right) \sin\theta}{\left(g^2+\ur ^2\right)^2} \right\}
\end{equation}
and for hopfions of the second type ($\chi=3\pi/2$) is
\begin{equation}
 \begin{aligned}
  \{\umIIr,\umIIt,\umIIf\} = &
  \biggl\{
  \frac{4 g^2 \ur ^2 \cos3\ut+\left(g^2-\ur ^2\right)^2 \cos\ut}{\left(g^2+\ur ^2\right)^2}, \\
  & - \sin\ut \left(1+\frac{8 g^2 \ur ^2 \cos2\ut}{\left(g^2+\ur^2\right)^2}\right),
  -\frac{4 g\ur\left(g^2-\ur^2\right) \sin\ut}{\left(g^2+\ur^2\right)^2}
  \biggr\},
 \end{aligned}
\end{equation}
where $g=1-f(\ur)$. Evidently (because the vector field components do not depend on the azimuthal angle $\uf$), the $\uHopf=1$ hopfions are axially symmetric. That is, their cross section by any plane, passing through the polar ($\uO\uZ$) coordinate system axis, contains exactly the same distribution of in-plane and out-of-plane components of the magnetization vector.

\section{Hopfion energy terms}
\label{sec:energyFuncs}
The exchange term of the functional can be calculated directly from its definition, performing the angular integration. It is the same for type~I and type~II hopfions:
\begin{equation}
 \uPex=\frac{64 \pi  \ur ^2 \left(3 g^2-2 g \ur  g' +\ur ^2 \left(g'\right)^2 \right)}{3 \left(g^2+\ur
   ^2\right)^2},
\end{equation}
where $g=g(\ur)=1-f(\ur)$ and $g'=g'(\ur)=-f'(\ur)$. This term depends on $\ur$, $f(\ur)$ and its derivative $f'(\ur)$.

The Zeeman and the anisotropy terms are also the same for both hopfion types and do not contain the derivative $f'(\ur)$:
\begin{eqnarray}
 \uPz & = & \frac{64 \pi  g^2 \ur ^4}{3 \left(g^2+\ur ^2\right)^2}, \\
 \uPa & = & \frac{128 \pi  g^2 \ur ^4 \left(5 g^4-6 g^2 \ur ^2+5 \ur ^4\right)}{15 \left(g^2+\ur ^2\right)^4}.
\end{eqnarray}

The Dzyaloshinskii-Moriya term can also be directly evaluated from the definition, but, unlike the previous terms, it is different for type~I and type~II hopfions
\begin{eqnarray}
\uPIdm & = & -\frac{32 \pi  \ur ^2 \left(-3 g^3+\ur  \left(g^2+\ur ^2\right) g'+g \ur ^2\right)}{3 \left(g^2+\ur
   ^2\right)^2} ,\\
\uPIIdm & = & \frac{32 \pi  \ur ^2 \left(-15 g^7+55 g^5 \ur ^2-53 g^3 \ur ^4+5 g \ur ^6\right)}{15 \left(g^2+\ur
   ^2\right)^4} + \nonumber \\
& & \frac{32 \pi  \ur ^3 \left(5 g^4-6 g^2 \ur ^2+5 \ur ^4\right) g'}{15 \left(g^2+\ur ^2\right)^3} .
\end{eqnarray}

To compute the magnetostatic self-energy, first, we need to evaluate the density of its volume magnetic charges. This density is simply (ignoring the sign, which is irrelevant for the present calculation) divergence of the magnetization vector field (multiplied by $R$ to make it dimensionless). For each of the hopfion types the density of charges is
\begin{eqnarray}
 \sigma^\mathrm{I} & = & \frac{16 g^2 \ur  \cos \theta}{\left(g^2+\ur ^2\right)^2},\\
 \sigma^\mathrm{II} & = & -\frac{16 g \ur \left(g^3+2\ur (\ur ^2-g^2)g'-3 g \ur ^2\right)\cos\theta}{\left(g^2+\ur
   ^2\right)^3} \nonumber \\
   & & \frac{32 g \ur ^2 \left(\left(\ur ^2-g^2\right) g'-2 g \ur \right) \cos^3\theta}{\left(g^2+\ur
   ^2\right)^3} .
\end{eqnarray}
The magnetostatic self-energy is then
\begin{equation}
 \label{eq:totalNormMS}
 \frac{\ueEMS}{\mu^2} = \int_0^1 \uPms \ud\ur= \frac{1}{8\pi}\uiiint_{|\uvr|<1} \ud^3\uvr\uiiint_{|\uvrp|<1}\ud^3\uvrp\frac{\sigma(\uvr)\sigma(\uvrp)}{|\uvr-\uvrp|},
\end{equation}
where the factor $1/2$ is already taken into account and the integration is performed in dimensionless units. To factor this integral we shall use the identity
\begin{equation}
\label{eq:sphericalHarmonics}
 \frac{1}{\left|\uvr -\uvr^\prime\right|}=4\pi\sum\limits_{l=0}^{\infty}\frac{1}{2l+1}\frac{\ur_\mathrm{min}^l}{\ur_\mathrm{max}^{l+1}}\sum\limits_{m=-\infty}^{\infty}Y_l^m(\ut,\uf)\overline{Y_l^m(\ut^\prime,\uf^\prime)},
\end{equation}
where the spherical coordinates $\uvr=\{\ur,\ut,\uf\}$, $\uvr^\prime=\{\ur^\prime,\ut^\prime,\uf^\prime\}$ are used, $\ur_\mathrm{min}=\min(\ur,\ur^\prime)$, $\ur_\mathrm{max}=\max(\ur,\ur^\prime)$, $Y_l^m(\ut,\uf)$ are the spherical harmonics and the overline denotes the complex conjugate. Note that for the type~I hopfions, the sum in~\eqref{eq:sphericalHarmonics} contains only the $l=1$ term, while for the hopfions of the type~II $l=1$ and $l=3$ terms. Performing the angular integration, the magnetostatic energy term can be represented as
\begingroup
\allowdisplaybreaks
\begin{align}
 \uPIms = & \frac{64 \pi  g^2 \rho \, W_1(\rho ,\{f\})}{9 \left(g^2+\rho ^2\right)^2}, \\
 W_1 = & \int_0^\ur\!\!\ud\ur\,\frac{16 g^2 \ur ^4}{\left(g^2+\ur ^2\right)^2} , \\
 \uPIIms = & -\frac{64 g \left(\left(g^2-\rho ^2\right) g'+2 g \rho \right) V_1(\rho ,\{f\})}{\left(g^2+\rho ^2\right)^3} - \nonumber \\
 & \frac{32 g \rho  \left(5 g^3+4 \rho  \left(\rho ^2-g^2\right) g'-3 g \rho ^2\right) V_2(\rho ,\{f\})}{3
   \left(g^2+\rho ^2\right)^3}, \\
 V_1 = & \int_0^\ur\!\!\ud\ur\,\frac{256 \pi  g \ur ^7 \left(\left(\ur^2-g^2\right) g'-2 g \ur \right)}{1225 \left(g^2+\ur
   ^2\right)^3} , \\
 V_2 = & \int_0^\ur\!\!\ud\ur\,\frac{32 \pi  g \ur ^4 \left(3 g \ur^2 - 5 g^3-4 \ur  \left(\ur ^2-g^2\right) g'\right)}{75
   \left(g^2+\ur ^2\right)^3}.
\end{align}
\endgroup
The total normalized magnetostatic energy~\eqref{eq:totalNormMS} depends only on the shape of the hopfion profile function $f(\ur)$. For a simple test case $f(\ur)=\ur^2$ the above expressions give $\ueIEMS/\mu^2=\num{0.854101633}$ and $\ueIIEMS/\mu^2=\num{0.774089652}$. These values were also verified by direct numerical 6-fold integration of~\eqref{eq:totalNormMS}. 


\begin{thebibliography}{38}%
\makeatletter
\providecommand \@ifxundefined [1]{%
 \@ifx{#1\undefined}
}%
\providecommand \@ifnum [1]{%
 \ifnum #1\expandafter \@firstoftwo
 \else \expandafter \@secondoftwo
 \fi
}%
\providecommand \@ifx [1]{%
 \ifx #1\expandafter \@firstoftwo
 \else \expandafter \@secondoftwo
 \fi
}%
\providecommand \natexlab [1]{#1}%
\providecommand \enquote  [1]{``#1''}%
\providecommand \bibnamefont  [1]{#1}%
\providecommand \bibfnamefont [1]{#1}%
\providecommand \citenamefont [1]{#1}%
\providecommand \href@noop [0]{\@secondoftwo}%
\providecommand \href [0]{\begingroup \@sanitize@url \@href}%
\providecommand \@href[1]{\@@startlink{#1}\@@href}%
\providecommand \@@href[1]{\endgroup#1\@@endlink}%
\providecommand \@sanitize@url [0]{\catcode `\\12\catcode `\$12\catcode
  `\&12\catcode `\#12\catcode `\^12\catcode `\_12\catcode `\%12\relax}%
\providecommand \@@startlink[1]{}%
\providecommand \@@endlink[0]{}%
\providecommand \url  [0]{\begingroup\@sanitize@url \@url }%
\providecommand \@url [1]{\endgroup\@href {#1}{\urlprefix }}%
\providecommand \urlprefix  [0]{URL }%
\providecommand \Eprint [0]{\href }%
\providecommand \doibase [0]{https://doi.org/}%
\providecommand \selectlanguage [0]{\@gobble}%
\providecommand \bibinfo  [0]{\@secondoftwo}%
\providecommand \bibfield  [0]{\@secondoftwo}%
\providecommand \translation [1]{[#1]}%
\providecommand \BibitemOpen [0]{}%
\providecommand \bibitemStop [0]{}%
\providecommand \bibitemNoStop [0]{.\EOS\space}%
\providecommand \EOS [0]{\spacefactor3000\relax}%
\providecommand \BibitemShut  [1]{\csname bibitem#1\endcsname}%
\let\auto@bib@innerbib\@empty
\bibitem [{\citenamefont {Thomson}(1869)}]{thomson1869}%
  \BibitemOpen
  \bibfield  {author} {\bibinfo {author} {\bibfnamefont {W.}~\bibnamefont
  {Thomson}},\ }\bibfield  {title} {\bibinfo {title} {{4. On Vortex Atoms}},\
  }\href {https://doi.org/10.1017/S0370164600045430} {\bibfield  {journal}
  {\bibinfo  {journal} {Proceedings of the Royal Society of Edinburgh}\
  }\textbf {\bibinfo {volume} {6}},\ \bibinfo {pages} {94–105} (\bibinfo
  {year} {1869})}\BibitemShut {NoStop}%
\bibitem [{\citenamefont {Crowell}\ and\ \citenamefont
  {Fox}(1977)}]{knot_theory_book_1977}%
  \BibitemOpen
  \bibfield  {author} {\bibinfo {author} {\bibfnamefont {R.~H.}\ \bibnamefont
  {Crowell}}\ and\ \bibinfo {author} {\bibfnamefont {R.~H.}\ \bibnamefont
  {Fox}},\ }\href {https://doi.org/10.1007/978-1-4612-9935-6} {\emph {\bibinfo
  {title} {{Introduction to Knot Theory}}}}\ (\bibinfo  {publisher} {Springer
  New York},\ \bibinfo {year} {1977})\BibitemShut {NoStop}%
\bibitem [{\citenamefont {Belavin}\ and\ \citenamefont
  {Polyakov}(1975)}]{BP75}%
  \BibitemOpen
  \bibfield  {author} {\bibinfo {author} {\bibfnamefont {A.~A.}\ \bibnamefont
  {Belavin}}\ and\ \bibinfo {author} {\bibfnamefont {A.~M.}\ \bibnamefont
  {Polyakov}},\ }\bibfield  {title} {\bibinfo {title} {{Metastable states of
  two-dimensional isotropic ferromagnet.}},\ }\href@noop {} {\bibfield
  {journal} {\bibinfo  {journal} {ZETP lett.}\ }\textbf {\bibinfo {volume}
  {22}},\ \bibinfo {pages} {245–247} (\bibinfo {year} {1975})}\BibitemShut
  {NoStop}%
\bibitem [{\citenamefont {Bobeck}\ and\ \citenamefont {{Della
  Torre}}(1975)}]{bobeck1975bubbles}%
  \BibitemOpen
  \bibfield  {author} {\bibinfo {author} {\bibfnamefont {A.~H.}\ \bibnamefont
  {Bobeck}}\ and\ \bibinfo {author} {\bibfnamefont {E.}~\bibnamefont {{Della
  Torre}}},\ }\href@noop {} {\emph {\bibinfo {title} {{Magnetic bubbles}}}}\
  (\bibinfo  {publisher} {North-Holland},\ \bibinfo {address} {Amsterdam},\
  \bibinfo {year} {1975})\BibitemShut {NoStop}%
\bibitem [{\citenamefont {Rößler}\ \emph {et~al.}(2006)\citenamefont
  {Rößler}, \citenamefont {Bogdanov},\ and\ \citenamefont
  {Pfleiderer}}]{RBP2006}%
  \BibitemOpen
  \bibfield  {author} {\bibinfo {author} {\bibfnamefont {U.~K.}\ \bibnamefont
  {Rößler}}, \bibinfo {author} {\bibfnamefont {A.~N.}\ \bibnamefont
  {Bogdanov}},\ and\ \bibinfo {author} {\bibfnamefont {C.}~\bibnamefont
  {Pfleiderer}},\ }\bibfield  {title} {\bibinfo {title} {{Spontaneous skyrmion
  ground states in magnetic metals}},\ }\href
  {https://doi.org/10.1038/nature05056} {\bibfield  {journal} {\bibinfo
  {journal} {Nature}\ }\textbf {\bibinfo {volume} {442}},\ \bibinfo {pages}
  {797–801} (\bibinfo {year} {2006})}\BibitemShut {NoStop}%
\bibitem [{\citenamefont {Usov}\ and\ \citenamefont {Peschany}(1993)}]{UP93}%
  \BibitemOpen
  \bibfield  {author} {\bibinfo {author} {\bibfnamefont {N.~A.}\ \bibnamefont
  {Usov}}\ and\ \bibinfo {author} {\bibfnamefont {S.~E.}\ \bibnamefont
  {Peschany}},\ }\bibfield  {title} {\bibinfo {title} {{Magnetization curling
  in a fine cylindrical particle}},\ }\href
  {https://doi.org/10.1016/0304-8853(93)90428-5} {\bibfield  {journal}
  {\bibinfo  {journal} {{J. Magn. Magn. Mater.}}\ }\textbf {\bibinfo {volume}
  {118}},\ \bibinfo {pages} {L290–L294} (\bibinfo {year} {1993})}\BibitemShut
  {NoStop}%
\bibitem [{\citenamefont {Metlov}(2010)}]{M10}%
  \BibitemOpen
  \bibfield  {author} {\bibinfo {author} {\bibfnamefont {K.~L.}\ \bibnamefont
  {Metlov}},\ }\bibfield  {title} {\bibinfo {title} {{Magnetization patterns in
  ferromagnetic nano-elements as functions of complex variable}},\ }\href
  {https://doi.org/10.1103/PhysRevLett.105.107201} {\bibfield  {journal}
  {\bibinfo  {journal} {Phys. Rev. Lett.}\ }\textbf {\bibinfo {volume} {105}},\
  \bibinfo {pages} {107201} (\bibinfo {year} {2010})}\BibitemShut {NoStop}%
\bibitem [{\citenamefont {Doring}(1968)}]{doring68}%
  \BibitemOpen
  \bibfield  {author} {\bibinfo {author} {\bibfnamefont {W.}~\bibnamefont
  {Doring}},\ }\bibfield  {title} {\bibinfo {title} {{Point Singularities in
  Micromagnetism}},\ }\href {https://doi.org/10.1063/1.1656144} {\bibfield
  {journal} {\bibinfo  {journal} {{J. Appl. Phys.}}\ }\textbf {\bibinfo
  {volume} {39}},\ \bibinfo {pages} {1006–1007} (\bibinfo {year}
  {1968})}\BibitemShut {NoStop}%
\bibitem [{\citenamefont {Faddeev}(1976)}]{Faddeev1976}%
  \BibitemOpen
  \bibfield  {author} {\bibinfo {author} {\bibfnamefont {L.~D.}\ \bibnamefont
  {Faddeev}},\ }\bibfield  {title} {\bibinfo {title} {{Some comments on the
  many-dimensional solitons}},\ }\href {https://doi.org/10.1007/BF00398483}
  {\bibfield  {journal} {\bibinfo  {journal} {Lett. Math. Phys.}\ }\textbf
  {\bibinfo {volume} {1}},\ \bibinfo {pages} {289–293} (\bibinfo {year}
  {1976})}\BibitemShut {NoStop}%
\bibitem [{\citenamefont {Hopf}(1931)}]{Hopf1931}%
  \BibitemOpen
  \bibfield  {author} {\bibinfo {author} {\bibfnamefont {H.}~\bibnamefont
  {Hopf}},\ }\bibfield  {title} {\bibinfo {title} {{Über die Abbildungen der
  dreidimensionalen Sphäre auf die Kugelfläche}},\ }\href
  {https://doi.org/10.1007/BF01457962} {\bibfield  {journal} {\bibinfo
  {journal} {Mathematische Annalen}\ }\textbf {\bibinfo {volume} {104}},\
  \bibinfo {pages} {637–665} (\bibinfo {year} {1931})}\BibitemShut {NoStop}%
\bibitem [{\citenamefont {Whitehead}(1947)}]{whitehead1947}%
  \BibitemOpen
  \bibfield  {author} {\bibinfo {author} {\bibfnamefont {J.~H.~C.}\
  \bibnamefont {Whitehead}},\ }\bibfield  {title} {\bibinfo {title} {{An
  Expression of Hopf's Invariant as an Integral}},\ }\href
  {https://doi.org/10.1073/pnas.33.5.117} {\bibfield  {journal} {\bibinfo
  {journal} {{Proc. Natl. Acad. Sci. U.S.A.}}\ }\textbf {\bibinfo {volume}
  {33}},\ \bibinfo {pages} {117–123} (\bibinfo {year} {1947})}\BibitemShut
  {NoStop}%
\bibitem [{\citenamefont {Nicole}(1978)}]{Nicole1978}%
  \BibitemOpen
  \bibfield  {author} {\bibinfo {author} {\bibfnamefont {D.~A.}\ \bibnamefont
  {Nicole}},\ }\bibfield  {title} {\bibinfo {title} {{Solitons with
  non-vanishing Hopf index}},\ }\href
  {https://doi.org/10.1088/0305-4616/4/9/008} {\bibfield  {journal} {\bibinfo
  {journal} {Journal of Physics G: Nuclear Physics}\ }\textbf {\bibinfo
  {volume} {4}},\ \bibinfo {pages} {1363–1369} (\bibinfo {year}
  {1978})}\BibitemShut {NoStop}%
\bibitem [{\citenamefont {Rybakov}\ \emph {et~al.}(2022)\citenamefont
  {Rybakov}, \citenamefont {Kiselev}, \citenamefont {Borisov}, \citenamefont
  {Döring}, \citenamefont {Melcher},\ and\ \citenamefont
  {Blügel}}]{RKBDMB2022}%
  \BibitemOpen
  \bibfield  {author} {\bibinfo {author} {\bibfnamefont {F.~N.}\ \bibnamefont
  {Rybakov}}, \bibinfo {author} {\bibfnamefont {N.~S.}\ \bibnamefont
  {Kiselev}}, \bibinfo {author} {\bibfnamefont {A.~B.}\ \bibnamefont
  {Borisov}}, \bibinfo {author} {\bibfnamefont {L.}~\bibnamefont {Döring}},
  \bibinfo {author} {\bibfnamefont {C.}~\bibnamefont {Melcher}},\ and\ \bibinfo
  {author} {\bibfnamefont {S.}~\bibnamefont {Blügel}},\ }\bibfield  {title}
  {\bibinfo {title} {{Magnetic hopfions in solids}},\ }\href
  {https://doi.org/10.1063/5.0099942} {\bibfield  {journal} {\bibinfo
  {journal} {APL Mater.}\ }\textbf {\bibinfo {volume} {10}},\ \bibinfo {pages}
  {111113} (\bibinfo {year} {2022})}\BibitemShut {NoStop}%
\bibitem [{\citenamefont {Sallermann}\ \emph {et~al.}(2023)\citenamefont
  {Sallermann}, \citenamefont {Jónsson},\ and\ \citenamefont
  {Blügel}}]{blugel2023}%
  \BibitemOpen
  \bibfield  {author} {\bibinfo {author} {\bibfnamefont {M.}~\bibnamefont
  {Sallermann}}, \bibinfo {author} {\bibfnamefont {H.}~\bibnamefont
  {Jónsson}},\ and\ \bibinfo {author} {\bibfnamefont {S.}~\bibnamefont
  {Blügel}},\ }\bibfield  {title} {\bibinfo {title} {{Stability of hopfions in
  bulk magnets with competing exchange interactions}},\ }\href
  {https://doi.org/10.1103/PhysRevB.107.104404} {\bibfield  {journal} {\bibinfo
   {journal} {Phys. Rev. B}\ }\textbf {\bibinfo {volume} {107}},\ \bibinfo
  {pages} {104404} (\bibinfo {year} {2023})}\BibitemShut {NoStop}%
\bibitem [{\citenamefont {Dzyloshinskii}\ and\ \citenamefont
  {Ivanov}(1979)}]{DI79}%
  \BibitemOpen
  \bibfield  {author} {\bibinfo {author} {\bibfnamefont {I.~E.}\ \bibnamefont
  {Dzyloshinskii}}\ and\ \bibinfo {author} {\bibfnamefont {B.~A.}\ \bibnamefont
  {Ivanov}},\ }\bibfield  {title} {\bibinfo {title} {{Localized topological
  solitons in a ferromagnet}},\ }\href@noop {} {\bibfield  {journal} {\bibinfo
  {journal} {JETP Lett.}\ }\textbf {\bibinfo {volume} {29}},\ \bibinfo {pages}
  {540–542} (\bibinfo {year} {1979})}\BibitemShut {NoStop}%
\bibitem [{\citenamefont {Kosevich}\ \emph {et~al.}(1990)\citenamefont
  {Kosevich}, \citenamefont {Ivanov},\ and\ \citenamefont {Kovalev}}]{KIK90}%
  \BibitemOpen
  \bibfield  {author} {\bibinfo {author} {\bibfnamefont {A.~M.}\ \bibnamefont
  {Kosevich}}, \bibinfo {author} {\bibfnamefont {B.~A.}\ \bibnamefont
  {Ivanov}},\ and\ \bibinfo {author} {\bibfnamefont {A.~S.}\ \bibnamefont
  {Kovalev}},\ }\bibfield  {title} {\bibinfo {title} {{Magnetic solitons}},\
  }\href {https://doi.org/10.1016/0370-1573(90)90130-T} {\bibfield  {journal}
  {\bibinfo  {journal} {Phys. Rep.}\ }\textbf {\bibinfo {volume} {194}},\
  \bibinfo {pages} {117–238} (\bibinfo {year} {1990})}\BibitemShut {NoStop}%
\bibitem [{\citenamefont {Borisov}\ and\ \citenamefont
  {Kiselev}(2011)}]{BK11-book-v2}%
  \BibitemOpen
  \bibfield  {author} {\bibinfo {author} {\bibfnamefont {A.~B.}\ \bibnamefont
  {Borisov}}\ and\ \bibinfo {author} {\bibfnamefont {V.}~\bibnamefont
  {Kiselev}},\ }\href@noop {} {\emph {\bibinfo {title} {{Topological Solitons,
  Two- and Three-Dimensional Patterns}}}},\ Vol.~\bibinfo {volume} {2}\
  (\bibinfo  {publisher} {{UB Press of RAS}},\ \bibinfo {address}
  {Ekaterinburg},\ \bibinfo {year} {2011})\BibitemShut {NoStop}%
\bibitem [{\citenamefont {Sutcliffe}(2018)}]{sutcliffe2018}%
  \BibitemOpen
  \bibfield  {author} {\bibinfo {author} {\bibfnamefont {P.}~\bibnamefont
  {Sutcliffe}},\ }\bibfield  {title} {\bibinfo {title} {{Hopfions in chiral
  magnets}},\ }\href {https://doi.org/10.1088/1751-8121/aad521} {\bibfield
  {journal} {\bibinfo  {journal} {J. Phys. A: Math. Theor.}\ }\textbf {\bibinfo
  {volume} {51}},\ \bibinfo {pages} {375401} (\bibinfo {year}
  {2018})}\BibitemShut {NoStop}%
\bibitem [{\citenamefont {Liu}\ \emph {et~al.}(2018)\citenamefont {Liu},
  \citenamefont {Lake},\ and\ \citenamefont {Zang}}]{lake2018}%
  \BibitemOpen
  \bibfield  {author} {\bibinfo {author} {\bibfnamefont {Y.}~\bibnamefont
  {Liu}}, \bibinfo {author} {\bibfnamefont {R.~K.}\ \bibnamefont {Lake}},\ and\
  \bibinfo {author} {\bibfnamefont {J.}~\bibnamefont {Zang}},\ }\bibfield
  {title} {\bibinfo {title} {{Binding a hopfion in a chiral magnet nanodisk}},\
  }\href {https://doi.org/10.1103/PhysRevB.98.174437} {\bibfield  {journal}
  {\bibinfo  {journal} {{Phys. Rev. B}}\ }\textbf {\bibinfo {volume} {98}},\
  \bibinfo {pages} {174437} (\bibinfo {year} {2018})}\BibitemShut {NoStop}%
\bibitem [{\citenamefont {Kent}\ \emph {et~al.}(2021)\citenamefont {Kent},
  \citenamefont {Reynolds}, \citenamefont {Raftrey}, \citenamefont {Campbell},
  \citenamefont {Virasawmy}, \citenamefont {Dhuey}, \citenamefont {Chopdekar},
  \citenamefont {Hierro-Rodriguez}, \citenamefont {Sorrentino}, \citenamefont
  {Pereiro}, \citenamefont {Ferrer}, \citenamefont {Hellman}, \citenamefont
  {Sutcliffe},\ and\ \citenamefont {Fischer}}]{kent2021}%
  \BibitemOpen
  \bibfield  {author} {\bibinfo {author} {\bibfnamefont {N.}~\bibnamefont
  {Kent}}, \bibinfo {author} {\bibfnamefont {N.}~\bibnamefont {Reynolds}},
  \bibinfo {author} {\bibfnamefont {D.}~\bibnamefont {Raftrey}}, \bibinfo
  {author} {\bibfnamefont {I.~T.~G.}\ \bibnamefont {Campbell}}, \bibinfo
  {author} {\bibfnamefont {S.}~\bibnamefont {Virasawmy}}, \bibinfo {author}
  {\bibfnamefont {S.}~\bibnamefont {Dhuey}}, \bibinfo {author} {\bibfnamefont
  {R.~V.}\ \bibnamefont {Chopdekar}}, \bibinfo {author} {\bibfnamefont
  {A.}~\bibnamefont {Hierro-Rodriguez}}, \bibinfo {author} {\bibfnamefont
  {A.}~\bibnamefont {Sorrentino}}, \bibinfo {author} {\bibfnamefont
  {E.}~\bibnamefont {Pereiro}}, \bibinfo {author} {\bibfnamefont
  {S.}~\bibnamefont {Ferrer}}, \bibinfo {author} {\bibfnamefont
  {F.}~\bibnamefont {Hellman}}, \bibinfo {author} {\bibfnamefont
  {P.}~\bibnamefont {Sutcliffe}},\ and\ \bibinfo {author} {\bibfnamefont
  {P.}~\bibnamefont {Fischer}},\ }\bibfield  {title} {\bibinfo {title}
  {{Creation and observation of Hopfions in magnetic multilayer systems}},\
  }\href {https://doi.org/10.1038/s41467-021-21846-5} {\bibfield  {journal}
  {\bibinfo  {journal} {Nat. Commun.}\ }\textbf {\bibinfo {volume} {12}},\
  \bibinfo {pages} {1562} (\bibinfo {year} {2021})}\BibitemShut {NoStop}%
\bibitem [{\citenamefont {Castillo-Sepúlveda}\ \emph
  {et~al.}(2023)\citenamefont {Castillo-Sepúlveda}, \citenamefont {Corona},
  \citenamefont {Saavedra}, \citenamefont {Laroze}, \citenamefont {Espejo},
  \citenamefont {Carvalho-Santos},\ and\ \citenamefont {Altbir}}]{altbir2023}%
  \BibitemOpen
  \bibfield  {author} {\bibinfo {author} {\bibfnamefont {S.}~\bibnamefont
  {Castillo-Sepúlveda}}, \bibinfo {author} {\bibfnamefont {R.~M.}\
  \bibnamefont {Corona}}, \bibinfo {author} {\bibfnamefont {E.}~\bibnamefont
  {Saavedra}}, \bibinfo {author} {\bibfnamefont {D.}~\bibnamefont {Laroze}},
  \bibinfo {author} {\bibfnamefont {A.~P.}\ \bibnamefont {Espejo}}, \bibinfo
  {author} {\bibfnamefont {V.~L.}\ \bibnamefont {Carvalho-Santos}},\ and\
  \bibinfo {author} {\bibfnamefont {D.}~\bibnamefont {Altbir}},\ }\bibfield
  {title} {\bibinfo {title} {{Nucleation and Stability of Toron Chains in
  Non-Centrosymmetric Magnetic Nanowires}},\ }\href
  {https://doi.org/10.3390/nano13121816} {\bibfield  {journal} {\bibinfo
  {journal} {Nanomaterials}\ }\textbf {\bibinfo {volume} {13}},\ \bibinfo
  {pages} {1816} (\bibinfo {year} {2023})}\BibitemShut {NoStop}%
\bibitem [{\citenamefont {Zheng}\ \emph {et~al.}(2023)\citenamefont {Zheng},
  \citenamefont {Kiselev}, \citenamefont {Rybakov}, \citenamefont {Yang},
  \citenamefont {Shi}, \citenamefont {Blügel},\ and\ \citenamefont
  {Dunin-Borkowski}}]{zheng2023}%
  \BibitemOpen
  \bibfield  {author} {\bibinfo {author} {\bibfnamefont {F.}~\bibnamefont
  {Zheng}}, \bibinfo {author} {\bibfnamefont {N.~S.}\ \bibnamefont {Kiselev}},
  \bibinfo {author} {\bibfnamefont {F.~N.}\ \bibnamefont {Rybakov}}, \bibinfo
  {author} {\bibfnamefont {L.}~\bibnamefont {Yang}}, \bibinfo {author}
  {\bibfnamefont {W.}~\bibnamefont {Shi}}, \bibinfo {author} {\bibfnamefont
  {S.}~\bibnamefont {Blügel}},\ and\ \bibinfo {author} {\bibfnamefont {R.~E.}\
  \bibnamefont {Dunin-Borkowski}},\ }\bibfield  {title} {\bibinfo {title}
  {{Hopfion rings in a cubic chiral magnet}},\ }\href
  {https://doi.org/10.1038/s41586-023-06658-5} {\bibfield  {journal} {\bibinfo
  {journal} {Nature}\ }\textbf {\bibinfo {volume} {623}},\ \bibinfo {pages}
  {718–723} (\bibinfo {year} {2023})}\BibitemShut {NoStop}%
\bibitem [{\citenamefont {Metlov}(2023)}]{M2023_TwoTypes}%
  \BibitemOpen
  \bibfield  {author} {\bibinfo {author} {\bibfnamefont {K.~L.}\ \bibnamefont
  {Metlov}},\ }\bibfield  {title} {\bibinfo {title} {{Two types of metastable
  hopfions in bulk magnets}},\ }\href
  {https://doi.org/10.1016/j.physd.2022.133561} {\bibfield  {journal} {\bibinfo
   {journal} {{Physica D}}\ }\textbf {\bibinfo {volume} {443}},\ \bibinfo
  {pages} {133561} (\bibinfo {year} {2023})}\BibitemShut {NoStop}%
\bibitem [{\citenamefont {Baryakhtar}\ and\ \citenamefont
  {Stefanovsky}(1970)}]{BS1970}%
  \BibitemOpen
  \bibfield  {author} {\bibinfo {author} {\bibfnamefont {V.~G.}\ \bibnamefont
  {Baryakhtar}}\ and\ \bibinfo {author} {\bibfnamefont {E.~P.}\ \bibnamefont
  {Stefanovsky}},\ }\bibfield  {title} {\bibinfo {title} {{Spin wave spectrum
  in antiferromagnets having a spiral magnetic structure}},\ }\href@noop {}
  {\bibfield  {journal} {\bibinfo  {journal} {Sov. Phys. Solid State}\ }\textbf
  {\bibinfo {volume} {11}},\ \bibinfo {pages} {1566–1572} (\bibinfo {year}
  {1970})}\BibitemShut {NoStop}%
\bibitem [{\citenamefont {Bak}\ and\ \citenamefont {Jensen}(1980)}]{BJ1980}%
  \BibitemOpen
  \bibfield  {author} {\bibinfo {author} {\bibfnamefont {P.}~\bibnamefont
  {Bak}}\ and\ \bibinfo {author} {\bibfnamefont {M.~H.}\ \bibnamefont
  {Jensen}},\ }\bibfield  {title} {\bibinfo {title} {{Theory of helical
  magnetic structures and phase transitions in {MnSi} and {FeGe}}},\ }\href
  {https://doi.org/10.1088/0022-3719/13/31/002} {\bibfield  {journal} {\bibinfo
   {journal} {{J. Phys. C}}\ }\textbf {\bibinfo {volume} {13}},\ \bibinfo
  {pages} {L881–L885} (\bibinfo {year} {1980})}\BibitemShut {NoStop}%
\bibitem [{\citenamefont {Aharoni}(1996)}]{AharoniBook}%
  \BibitemOpen
  \bibfield  {author} {\bibinfo {author} {\bibfnamefont {A.}~\bibnamefont
  {Aharoni}},\ }\href@noop {} {\emph {\bibinfo {title} {{Introduction to the
  theory of ferromagnetism}}}}\ (\bibinfo  {publisher} {Oxford University
  Press},\ \bibinfo {address} {Oxford},\ \bibinfo {year} {1996})\BibitemShut
  {NoStop}%
\bibitem [{\citenamefont {Garst}\ \emph {et~al.}(2017)\citenamefont {Garst},
  \citenamefont {Waizner},\ and\ \citenamefont {Grundler}}]{GWG2017}%
  \BibitemOpen
  \bibfield  {author} {\bibinfo {author} {\bibfnamefont {M.}~\bibnamefont
  {Garst}}, \bibinfo {author} {\bibfnamefont {J.}~\bibnamefont {Waizner}},\
  and\ \bibinfo {author} {\bibfnamefont {D.}~\bibnamefont {Grundler}},\
  }\bibfield  {title} {\bibinfo {title} {{Collective spin excitations of
  helices and magnetic skyrmions: review and perspectives of magnonics in
  non-centrosymmetric magnets}},\ }\href
  {https://doi.org/10.1088/1361-6463/aa7573} {\bibfield  {journal} {\bibinfo
  {journal} {{J. Phys. D}}\ }\textbf {\bibinfo {volume} {50}},\ \bibinfo
  {pages} {293002} (\bibinfo {year} {2017})}\BibitemShut {NoStop}%
\bibitem [{\citenamefont {Mattheij}\ and\ \citenamefont
  {Staarink}(1984)}]{MUS1984}%
  \BibitemOpen
  \bibfield  {author} {\bibinfo {author} {\bibfnamefont {R.~M.~M.}\
  \bibnamefont {Mattheij}}\ and\ \bibinfo {author} {\bibfnamefont {G.~W.~M.}\
  \bibnamefont {Staarink}},\ }\bibfield  {title} {\bibinfo {title} {{An
  Efficient Algorithm for Solving General Linear Two-Point BVP}},\ }\href
  {https://doi.org/10.1137/0905053} {\bibfield  {journal} {\bibinfo  {journal}
  {SIAM Journal on Scientific and Statistical Computing}\ }\textbf {\bibinfo
  {volume} {5}},\ \bibinfo {pages} {745–763} (\bibinfo {year}
  {1984})}\BibitemShut {NoStop}%
\bibitem [{\citenamefont {Metlov}(2024{\natexlab{a}})}]{magnhopf}%
  \BibitemOpen
  \bibfield  {author} {\bibinfo {author} {\bibfnamefont {K.~L.}\ \bibnamefont
  {Metlov}},\ }\href@noop {} {\bibinfo {title} {{Source code}}},\ \bibinfo
  {howpublished} {\url{https://github.com/metlov/magnhopf}} (\bibinfo {year}
  {2024}{\natexlab{a}})\BibitemShut {NoStop}%
\bibitem [{\citenamefont {Bogdanov}\ and\ \citenamefont
  {Yablonskii}(1989)}]{BY1989}%
  \BibitemOpen
  \bibfield  {author} {\bibinfo {author} {\bibfnamefont {A.}~\bibnamefont
  {Bogdanov}}\ and\ \bibinfo {author} {\bibfnamefont {D.}~\bibnamefont
  {Yablonskii}},\ }\bibfield  {title} {\bibinfo {title} {{Thermodynamically
  stable {"}vortices{"} in magnetically ordered crystals. The mixed state of
  magnets}},\ }\href@noop {} {\bibfield  {journal} {\bibinfo  {journal} {Sov.
  Phys. JETP}\ }\textbf {\bibinfo {volume} {68}},\ \bibinfo {pages} {101}
  (\bibinfo {year} {1989})}\BibitemShut {NoStop}%
\bibitem [{\citenamefont {Bogdanov}\ and\ \citenamefont
  {Hubert}(1994)}]{BH1994}%
  \BibitemOpen
  \bibfield  {author} {\bibinfo {author} {\bibfnamefont {A.}~\bibnamefont
  {Bogdanov}}\ and\ \bibinfo {author} {\bibfnamefont {A.}~\bibnamefont
  {Hubert}},\ }\bibfield  {title} {\bibinfo {title} {{Thermodynamically stable
  magnetic vortex states in magnetic crystals}},\ }\href
  {https://doi.org/10.1016/0304-8853(94)90046-9} {\bibfield  {journal}
  {\bibinfo  {journal} {{J. Magn. Magn. Mater.}}\ }\textbf {\bibinfo {volume}
  {138}},\ \bibinfo {pages} {255–269} (\bibinfo {year} {1994})}\BibitemShut
  {NoStop}%
\bibitem [{\citenamefont {Hobart}(1963)}]{Hobart1963}%
  \BibitemOpen
  \bibfield  {author} {\bibinfo {author} {\bibfnamefont {R.~H.}\ \bibnamefont
  {Hobart}},\ }\bibfield  {title} {\bibinfo {title} {{On the Instability of a
  Class of Unitary Field Models}},\ }\href
  {https://doi.org/10.1088/0370-1328/82/2/306} {\bibfield  {journal} {\bibinfo
  {journal} {Proceedings of the Physical Society}\ }\textbf {\bibinfo {volume}
  {82}},\ \bibinfo {pages} {201} (\bibinfo {year} {1963})}\BibitemShut
  {NoStop}%
\bibitem [{\citenamefont {Derrick}(1964)}]{Derrick1964}%
  \BibitemOpen
  \bibfield  {author} {\bibinfo {author} {\bibfnamefont {G.~H.}\ \bibnamefont
  {Derrick}},\ }\bibfield  {title} {\bibinfo {title} {{Comments on Nonlinear
  Wave Equations as Models for Elementary Particles}},\ }\href
  {https://doi.org/10.1063/1.1704233} {\bibfield  {journal} {\bibinfo
  {journal} {{J. Math. Phys.}}\ }\textbf {\bibinfo {volume} {5}},\ \bibinfo
  {pages} {1252–1254} (\bibinfo {year} {1964})}\BibitemShut {NoStop}%
\bibitem [{\citenamefont {Metlov}(2001)}]{M01_CT}%
  \BibitemOpen
  \bibfield  {author} {\bibinfo {author} {\bibfnamefont {K.~L.}\ \bibnamefont
  {Metlov}},\ }\bibfield  {title} {\bibinfo {title} {{Simple analytical
  description of the cross-tie domain wall structure}},\ }\href
  {https://doi.org/10.1063/1.1409946} {\bibfield  {journal} {\bibinfo
  {journal} {{Appl. Phys. Lett.}}\ }\textbf {\bibinfo {volume} {79}},\ \bibinfo
  {pages} {2609–2611} (\bibinfo {year} {2001})}\BibitemShut {NoStop}%
\bibitem [{\citenamefont {Donnelly}\ \emph {et~al.}(2021)\citenamefont
  {Donnelly}, \citenamefont {Metlov}, \citenamefont {Scagnoli}, \citenamefont
  {Guizar-Sicairos}, \citenamefont {Holler}, \citenamefont {Bingham},
  \citenamefont {Raabe}, \citenamefont {Heyderman}, \citenamefont {Cooper},\
  and\ \citenamefont {Gliga}}]{CMSGHBRHCG2020}%
  \BibitemOpen
  \bibfield  {author} {\bibinfo {author} {\bibfnamefont {C.}~\bibnamefont
  {Donnelly}}, \bibinfo {author} {\bibfnamefont {K.~L.}\ \bibnamefont
  {Metlov}}, \bibinfo {author} {\bibfnamefont {V.}~\bibnamefont {Scagnoli}},
  \bibinfo {author} {\bibfnamefont {M.}~\bibnamefont {Guizar-Sicairos}},
  \bibinfo {author} {\bibfnamefont {M.}~\bibnamefont {Holler}}, \bibinfo
  {author} {\bibfnamefont {N.~S.}\ \bibnamefont {Bingham}}, \bibinfo {author}
  {\bibfnamefont {J.}~\bibnamefont {Raabe}}, \bibinfo {author} {\bibfnamefont
  {L.~J.}\ \bibnamefont {Heyderman}}, \bibinfo {author} {\bibfnamefont
  {N.}~\bibnamefont {Cooper}},\ and\ \bibinfo {author} {\bibfnamefont
  {S.}~\bibnamefont {Gliga}},\ }\bibfield  {title} {\bibinfo {title}
  {{Experimental Observation of Vortex Rings in a Bulk Magnet}},\ }\href
  {https://doi.org/10.1038/s41567-020-01057-3} {\bibfield  {journal} {\bibinfo
  {journal} {{Nat. Phys.}}\ }\textbf {\bibinfo {volume} {17}},\ \bibinfo
  {pages} {316–321} (\bibinfo {year} {2021})},\ \Eprint
  {https://arxiv.org/abs/arXiv:2009.04226} {arXiv:2009.04226} \BibitemShut
  {NoStop}%
\bibitem [{\citenamefont {Tai}\ and\ \citenamefont {Smalyukh}(2018)}]{TS2018}%
  \BibitemOpen
  \bibfield  {author} {\bibinfo {author} {\bibfnamefont {J.-S.~B.}\
  \bibnamefont {Tai}}\ and\ \bibinfo {author} {\bibfnamefont {I.~I.}\
  \bibnamefont {Smalyukh}},\ }\bibfield  {title} {\bibinfo {title} {{Static
  Hopf Solitons and Knotted Emergent Fields in Solid-State Noncentrosymmetric
  Magnetic Nanostructures}},\ }\href
  {https://doi.org/10.1103/PhysRevLett.121.187201} {\bibfield  {journal}
  {\bibinfo  {journal} {Phys. Rev. Lett.}\ }\textbf {\bibinfo {volume} {121}},\
  \bibinfo {pages} {187201} (\bibinfo {year} {2018})}\BibitemShut {NoStop}%
\bibitem [{\citenamefont {Metlov}(2024{\natexlab{b}})}]{M2024Kourovka}%
  \BibitemOpen
  \bibfield  {author} {\bibinfo {author} {\bibfnamefont {K.~L.}\ \bibnamefont
  {Metlov}},\ }\bibfield  {title} {\bibinfo {title} {{Two types of metastable
  magnetic hopfions and their elliptical stability}}} (\bibinfo {year}
  {2024}{\natexlab{b}}),\ \bibinfo {note} {oral presentation (in Russian) at XL
  International Winter School for Theoretical Physics “Kourovka”, 2-9
  February 2024, ski resort “Abzakovo”, Republic of Bashkortostan, Russian
  Federation}\BibitemShut {NoStop}%
\bibitem [{\citenamefont {{\ifmmode \check{S}\else Š\fi{}mejkal}}\ \emph
  {et~al.}(2022)\citenamefont {{\ifmmode \check{S}\else Š\fi{}mejkal}},
  \citenamefont {Sinova},\ and\ \citenamefont {Jungwirth}}]{SSJ2022}%
  \BibitemOpen
  \bibfield  {author} {\bibinfo {author} {\bibfnamefont {L.}~\bibnamefont
  {{\ifmmode \check{S}\else Š\fi{}mejkal}}}, \bibinfo {author} {\bibfnamefont
  {J.}~\bibnamefont {Sinova}},\ and\ \bibinfo {author} {\bibfnamefont
  {T.}~\bibnamefont {Jungwirth}},\ }\bibfield  {title} {\bibinfo {title}
  {{Emerging Research Landscape of Altermagnetism}},\ }\href
  {https://doi.org/10.1103/PhysRevX.12.040501} {\bibfield  {journal} {\bibinfo
  {journal} {Phys. Rev. X}\ }\textbf {\bibinfo {volume} {12}},\ \bibinfo
  {pages} {040501} (\bibinfo {year} {2022})}\BibitemShut {NoStop}%
\end{thebibliography}
%
\end{document}